\documentclass[sigconf]{acmart}
\usepackage{float}
\AtBeginDocument{%
  \providecommand\BibTeX{{%
    \normalfont B\kern-0.5em{\scshape i\kern-0.25em b}\kern-0.8em\TeX}}}
\copyrightyear{2024}
\acmYear{2024}
\setcopyright{rightsretained}
\acmConference[CAIN 2024]{Conference on AI Engineering Software Engineering for AI}{April 14--15, 2024}{Lisbon, Portugal}
\acmBooktitle{Conference on AI Engineering Software Engineering for AI (CAIN 2024), April 14--15, 2024, Lisbon, Portugal}\acmDOI{10.1145/3644815.3644981}
\acmISBN{979-8-4007-0591-5/24/04}
\begin{document}

\title{Automating Patch Set Generation from Code Review Comments Using Large Language Models}

%\author{Tajmilur Rahman, Rahul Singh, Yousuf Sultan}
\author{Tajmilur Rahman, Rahul Singh, Mir Yousuf Sultan}
\email{{rahman007, rahul041, mir002} @gannon.edu}
%\orcid{0000-0001-9629-8144}
\affiliation{%
  \institution{Gannon University}
  \streetaddress{109 University Square}
  \city{Erie}
  \state{PA}
  \country{USA}
  \postcode{16541}
}

\renewcommand{\shortauthors}{Tajmilur Rahman}

\begin{abstract}
The advent of Large Language Models (LLMs) has revolutionized various domains of artificial intelligence, including the realm of software engineering. 
In this research, we evaluate the efficacy of pre-trained LLMs in replicating the tasks traditionally performed by developers in response to code review comments. 
We provide code contexts to five popular LLMs and obtain the suggested code-changes (patch sets) derived from real-world code-review comments. 
The performance of each model is meticulously assessed by comparing their generated patch sets against the historical data of human-generated patch-sets from the same repositories. 
This comparative analysis aims to determine the accuracy, relevance, and depth of the LLMs' feedback, thereby evaluating their readiness to support developers in responding to code-review comments. \textit{\textbf{Novelty:}} This particular research area is still immature requiring a substantial amount of studies yet to be done. 
No prior research has compared the performance of existing Large Language Models (LLMs) in code-review comments. This in-progress study assesses current LLMs in code review and paves the way for future advancements in automated code quality assurance, reducing context-switching overhead due to interruptions from code change requests.
\end{abstract}

%\received{15 December 2023}
%\received[revised]{15 January 2024}
%\received[accepted]{5 February 2024}

\maketitle

\begin{figure*}[t!]
  \centering
  \includegraphics[width=0.8\textwidth]{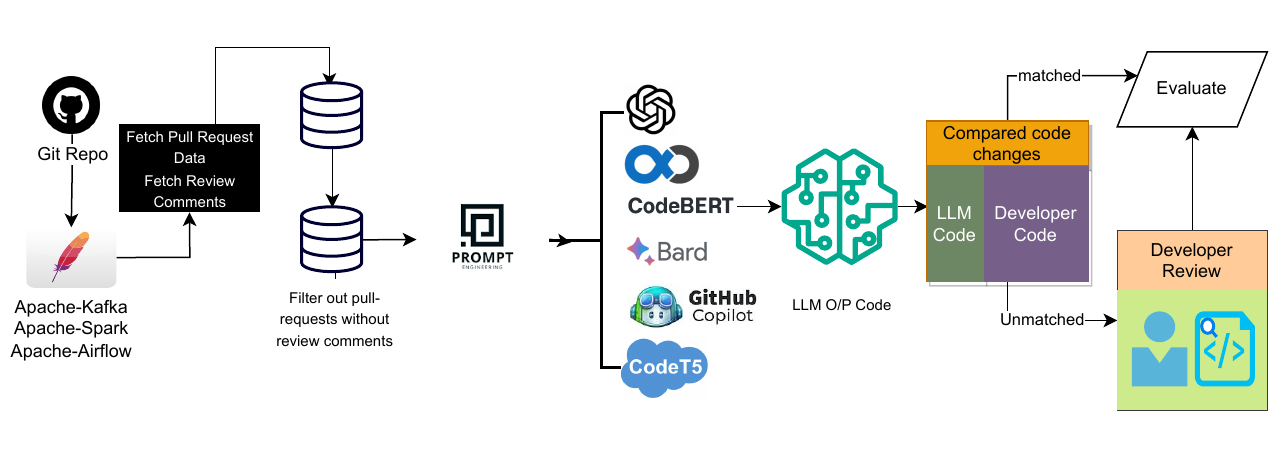}
  \caption{Study Design}
  \label{fig:method}
\end{figure*}

\section{Introduction}
Code review stands as a critical quality assurance practice, ensuring code correctness, maintainability, and adherence to coding standards. 
%Traditionally, this process has been the responsibility of human reviewers, whose expertise and experience guide the assessment of code contributions. 
%Developers often switch tasks immediately after completing a pull request. 
%However, due to the pressure of fast-paced development, there is a growing need for automated assistance in the code review process. 
Developers frequently transition to other tasks right after finalizing a pull request. This shift is driven by the demands of fast-paced development, where there is an increasing need for automated assistance in the code review process.
Frequent context switches caused by pull-request review comments can slow down developers. 
Recent advances in Artificial Intelligence, especially in Large Language Models (LLMs), present opportunities to automate aspects of code review, enhancing the efficiency of human reviewers~\cite{smith2020codereview}.

This research investigates the feasibility of employing pre-trained LLMs, such as OpenAI's GPT-4, Google's BARD, Microsoft's CodeBERT, Deep Mind's AlphaCode, and Salesforce's CodeT5, and GitHub Copilot in the context of automating code review. 
%These models, originally designed for natural language processing tasks, have shown remarkable capabilities in understanding and generating human-like text, which can be leveraged to analyze and critique 
The models were initially designed for natural language processing tasks. They have demonstrated remarkable capabilities in understanding and generating human-like text. These capabilities can be leveraged to analyze and critique code~\cite{devlin2018bert, brown2020language}. 
By presenting these models with code context and patch sets from real-world Pull Requests, we aim to explore the extent to which LLMs can replicate the nuanced evaluations typically performed by human reviewers.

In our approach, the performance of the LLMs will be rigorously assessed by comparing their understanding ability of code-review comments and code-change suggestions against the historical data of human code reviews from existing repositories. 
Metrics such as accuracy, relevance, and actionability of feedback will serve as benchmarks for evaluation~\cite{wang2019learning}. 

\subsection{Research Questions}
Our study is structured around the following research questions aimed at comprehending how various Language Models (LLMs) respond to distinct code-review comments. The objective is to gain insights into the diverse ways LLMs react to different types of code-review feedback and compare the performance of today's LLMs.
\begin{enumerate}
    \item RQ1: How accurately can today's LLMs interpret and understand code-review comments?
    \item RQ2: Can today's LLMs generate necessary code-changes based on code-review comments?
    \item RQ3: How acceptable are the generated code-changes to the human developers given the variations of contexts in different software projects and what is the acceptance threshold?
    \item RQ4: How much contextual information must be embedded within a code-review comment to achieve the minimum acceptance threshold?
\end{enumerate}

\section{Related Works}
Tufano et al.~\cite{tufano2021towards} partially automated the manual code review process to reduce the time spent by developers ~\cite{Shuai2021CodeXGLUE}. 
They created two distinct deep-learning architectures. 
The first model learns code changes made by developers offering contributors a revised version of their code with recommended transformations before submission. 
The second model assists reviewers by automatically implementing their comments expressed in natural language.
While the performance of the models was promising, with the contributor-side model replicating code transformations for up to 16\% 
of cases and the reviewer-side model correctly implementing comments for up to 31\% of cases, there are still plenty of areas for improvement that require further study.
%Although, the performance of the models was promising, while the contributor-side model replicates code transformations for up to 16\% of cases, 
%and the reviewer-side model correctly implements comments for up to 31\% of cases.
%Still, there are plenty of rooms of improvement that require more studies. 

Palvannan and Brown~\cite{palvannan2023suggestion} explored the integration of a bot named ``SUGGESTION BOT'' 
into the software development cycle to enhance the efficiency of the peer code review process. 
The bot leverages GitHub's suggested changes functionality to automatically review code, 
aiming to address challenges faced by developers with non-comprehensive feedback and 
disruptive notifications from existing bots. 
%The researchers conducted a comparative empirical investigation, evaluating the SUGGESTION BOT's impact on the turnaround time of pull requests. 
%The study assesses the clarity and usefulness of comments provided by the bot and the results offer insights into the design of future systems and the enhancement of human-bot interactions in code review processes.

Li et al.~\cite{li2022automating} studied LLMs at a large scale that addresses the 
time-consuming nature of manual code reviews in software development by proposing an automated approach. 
The study focuses on leveraging pre-training techniques and introduces a model called ``CodeReviewer'' which acts as a code reviewer. 
%The researchers collected a substantial dataset from real-world code changes and reviews across nine popular programming languages. 
%CodeReviewer is evaluated on three key tasks related to code review activities: 
%code change quality estimation, review comment generation, and code refinement. 
%To assess performance, the researchers establish a high-quality benchmark dataset and conduct comprehensive experiments. 
The results indicate the effectiveness of CodeReviewer in automating various aspects of the code review process.

Compared to the existing studies, our study aims to develop a model to generate code change suggestions in response to code-review comments.
Since code-review is a way of transferring system knowledge to the newer team members, our aim is not to introduce a code reviewer bot to eliminate human involvement from the entire code-review process.
Instead, we would like to save the additional time that it takes for a developer to address the code-review comments.

\section{Methodology}
Figure~\ref{fig:method} shows the study design at a glance. 
We started with downloading 30K pull requests from three Apache projects: Kafka, Spark, and Airflow using GitHub APIs. 
We removed the pull requests that didn't have any review comments since we are interested in generating code changes based on code-review comments only.

We created various prompts to train five distinct language models: GPT-4.0, CodeBERT, BARD, Copilot, and CodeT5. 
After standardizing the prompt format across all models, the code, along with corresponding comments and prompts, is individually submitted to each Language Model (LLM). 
The generated outputs are then compared to existing code changes in repositories. 
If the suggested code change review matches existing repository data by 80\% or more ~\cite{Khalek80percent}, it is automatically integrated into the evaluation phase. 
Code suggestions matched below the 80\% threshold are directed to developers for manual review. 
Based on developers' feedback if a patch-set doesn't pass then that is counted as a failing code suggestion. 
However, if human reviewers approve the code-suggestion as an acceptable change to address the corresponding code-review, it goes to the evaluation phase. 
In the evaluation phase, the LLM models are evaluated based on the number of code-suggestions accepted from a model.

\bibliographystyle{ACM-Reference-Format}
\bibliography{main}

\end{document}